 \documentclass[final,3p,times,twocolumn]{elsarticle}
\usepackage{amssymb}
\usepackage{amsthm}

\usepackage{comment}
\usepackage{amsfonts}
\usepackage{amsmath}
\newtheorem{theorem}{Theorem}
\newtheorem{definition}{Definition}
\newtheorem{lemma}{Lemma}

\usepackage[ruled,vlined,linesnumbered,longend]{algorithm2e}
\usepackage{caption}
\usepackage{subcaption}
\usepackage{enumerate}
\usepackage{url}
\usepackage{balance}

\journal{Computer Networks}

\begin{document}

\begin{frontmatter}



\title{Radiation-constrained Algorithms for Wireless Energy Transfer in Ad hoc Networks\tnoteref{t1}}
\tnotetext[t1]{A preliminary version of this paper appeared in \cite{7164906}.}


\author{Sotiris Nikoletseas, Theofanis P. Raptis, Christoforos Raptopoulos}

\address{Department of Computer Engineering and Informatics, University of Patras, Greece}
\address{Computer Technology Institute and Press ``Diophantus'', Greece}

\begin{abstract}
We study the problem of efficiently charging a set of rechargeable nodes using a set of wireless chargers, under safety constraints on the electromagnetic radiation incurred. In particular, we define a new charging model that greatly differs from existing models in that it takes into account real technology restrictions of the chargers and nodes of the network, mainly regarding energy limitations. Our model also introduces non-linear constraints (in the time domain), that radically change the nature of the computational problems we consider. In this charging model, we present and study the \emph{Low Radiation Efficient Charging Problem} (LREC), in which we wish to optimize the amount of ``useful'' energy transferred from chargers to nodes (under constraints on the maximum level of imposed radiation). We present several fundamental properties of this problem and provide indications of its hardness. Finally, we propose an iterative local improvement heuristic for LREC, which runs in polynomial time and we evaluate its performance via simulation. Our algorithm decouples the computation of the objective function from the computation of the maximum radiation and also does not depend on the exact formula used for the computation of the electromagnetic radiation in each point of the network, achieving good trade-offs between charging efficiency and radiation control; it also exhibits good energy balance properties. We provide extensive simulation results supporting our claims and theoretical results.
\end{abstract}

\begin{keyword}
Wireless Energy Transfer \sep Ad hoc Networks \sep Electromagnetic Radiation \sep Algorithms


\end{keyword}

\end{frontmatter}


\section{Introduction}

Recent advances in embedded systems and the high adoption rates of truly portable, hand-held devices by the general public have motivated significant research efforts in \emph{Ad hoc Networks}. Due to the intrinsic mobile nature of these devices and the constraints characterizing them, the efficient management of energy has been one of the most prominent research targets. Energy in Ad hoc Networks is highly correlated to important operational aspects, such as network lifetime and resilience, quality of service-energy efficiency trade-offs and network throughput. Late technological advances in the domain of \emph{Wireless Energy Transfer} (WET) pave the way for novel methods for energy management in Ad hoc Networks and recent research efforts have already started considering network models that take into account these new technologies. 

In particular, the technology of highly-efficient Wireless Energy Transfer was proposed for efficient energy transmission over mid-range distances. The work in \cite{KKMJFS2007} has shown that through strongly coupled magnetic resonances, the efficiency of transferring 60 Watts of power over a distance in excess of 2 meters is as high as 40\%. Industry research also demonstrated that it is possible to improve transferring 60 Watts of power over a distance of up to 1 meter with efficiency of 75\% \cite{intel}. At present, commercial products utilizing Wireless Energy Transfer have been available on the market such as those in \cite{powercast}, \cite{murata} and \cite{ti}. Finally, the Wireless Power Consortium \cite{wpc}, with members including IC manufacturers, smartphone makers and telecom operators, was established to set the international standards for interoperable wireless charging. 

%The advantages of using WET technologies in Ad hoc Networks are several: At first, the highly constrained resource of energy can be managed in great detail and efficiency. Secondly, energy management can be performed passively from the perspective of the nodes, thus rendering obsolete the computational and communicational overhead introduced by complex energy management algorithms. Finally, the use of WET in Ad hoc Networks allows energy management to be studied and designed independently of the underlying network structure. 

However, the beneficial use of WET in Ad hoc Networks comes at a price with regards to real life applications. Wireless Energy Transfer introduces a new source of \emph{electromagnetic radiation} (EMR) that will co-exist with several other wireless technologies (i.e., Wi-Fi, Bluetooth, etc.). %This constantly enriching wireless environment has yet to be proven harmful from an epidemiological point of view. 
Exposure to high electromagnetic radiation, has been widely recognized as a \emph{threat to human health}. Its potential risks include but is not limited to mental diseases \cite{biology}, tissue impairment \cite{implants} and brain tumor \cite{oncology}. In addition, there has been solid evidence that pregnant women and children are even more vulnerable to high electromagnetic radiation exposure \cite{embryos1}, \cite{embryos2}. We note that particularly the radiation levels created by wireless power can he quite high, due to the strength of the electromagnetic fields created. Even if the impact of electromagnetic radiation can be considered controversial we believe it is worth understanding and control, without however compromising the quality of service offered to the user of wireless communications. For such systems, the broader aim would be to come up with radiation awareness in an adaptive manner, by providing design principles and studying key algorithmic and networking aspects of radiation aware wireless networking.  The Computer Science research community has already demonstrated relevant interest from an ICT perspective by considering restrictions in the amount of emitted EMR. This creates a new topic in algorithmic network design for Ad hoc Networks.

\subsection{Related Work}

%In the field of Wireless Energy Transfer, various problems in wireless networking settings have been investigated. Most of them investigate the efficient available energy management. 
Research efforts in Ad hoc Networks have already started considering network models that take into account WET technologies. For instance, wireless rechargeable sensor networks consist of sensor nodes, as well as few nodes with high energy supplies (wireless chargers). The latter are capable of fast charging sensor nodes, by using Wireless Energy Transfer technologies. In \cite{7296607} and \cite{comnet14}, the authors assume special mobile charging entities, which traverse the network and wirelessly replenish the energy of sensor nodes. Their methods are distributed, adaptive and perform well in detailed experimental simulations. %After detailed simulations in uniform and non-uniform network deployments, using three different underlying routing protocol families, the charging protocols' performance gets quite close to the performance of a global knowledge powerful method. 
In \cite{Madhja201589} and \cite{7164900} the authors employ chargers in sensor networks and collaboratively compute the coordination and charging processes. The authors also provide %limited network knowledge protocols, centralized global network knowledge protocols and 
protocols that grant the chargers the ability to perform multi-hop charging. In \cite{6990384}, the authors propose joint charging and rate allocation scheme that maximizes the network utility while satisfying the network sustainability requirement. The scheme is designed based on the observation that the energy repository of a sensor node is co-affected by specific factors. In \cite{Li11} a practical and efficient joint routing and charging scheme is proposed. %Through proactively guiding the routing activities in the network and delivering energy to where it is needed, this scheme replenishes energy into the network and improves the network energy utilization, thus prolonging the network lifetime.  
In \cite{julia}, the authors define a weighting function which evaluates several network parameters in order to prioritize the nodes during the charging process and based on this function they define three traversal strategies. In \cite{pimrc13}, the authors formulate a set of power flow problems and propose algorithms to solve them based on a detailed analysis on the problem structure. Moreover the authors further investigate the joint data and power flow problems. In \cite{6747302}, the authors propose a framework of joint wireless energy replenishment and anchor-point based mobile data gathering in sensor networks by considering various sources of energy consumption and time-varying nature of energy replenishment. 
%Applying the proposed algorithms, the authors obtain extensive numerical results on the optimal configurations of power flow and joint data/power flows, which yield instructive insights on practical system constructions.

The attention of researchers from many diverse research fields has been drawn in the field of electromagnetic radiation impact. Consequently, there has also been research on radiation related problems in the Ad hoc Networks context. In \cite{dcoss12}, the authors study the problem of electromagnetic radiation in wireless sensor networks and more specifically maintaining low radiation trajectories for a person moving in a sensor network area. the authors evaluate, mathematically the radiation in well known sensor network topologies and random geometric graphs. Then the authors implement online protocols and comparatively study their performance via simulation. Those heuristics achieve low radiation paths which are even close to an off-line optimum. In \cite{mobiwac12}, the authors focus on the problem of efficient data propagation in wireless sensor networks, trying to keep latency low while maintaining at low levels the radiation cumulated by wireless transmissions. The authors first propose greedy and oblivious routing heuristics that are radiation aware. They then combine them with temporal back-off schemes that use local properties of the network in order to spread radiation in a spatio-temporal way. The proposed radiation aware routing heuristics succeed to keep radiation levels low, while not increasing latency. In \cite{kranakis}, the authors consider the problem of covering a planar region, which includes a collection of buildings, with a minimum number of stations so that every point in the region is within the reach of a station, while at the same time no building is within the dangerous range of a station. However, those approaches are oriented towards network devices radiation, not addressing wireless chargers.

Some limited research has also been conducted in the cross-section of Wireless Energy Transfer and electromagnetic radiation in networking settings. In \cite{infocom14}, the authors study the problem of scheduling stationary chargers so that more energy can be received while no location in the field has electromagnetic radiation (EMR) exceeding a given threshold. The authors design a method that transfers the problem to two traditional problems, namely a multidimensional $0/1$ knapsack problem and a Fermat-Weber problem. The method includes constraint conversion and reduction, bounded EMR function approximation, area discretization and expansion, and a tailored Fermat-Weber algorithm. In order to evaluate the performance of their method, the authors build a testbed composed of 8 chargers. In \cite{7524385}, the authors propose a wireless charger placement scheme that guarantees EMR Safety for every location on the plane. In \cite{icdcs14}, the authors consider the problem of of scheduling stationary chargers with adjustable power, namely how to adjust the power of chargers so as to maximize the charging utility of the devices, while assuring that EMR intensity at any location in the field does not exceed a given threshold. The authors present an area discretization technique to help re-formulating the problem into a traditional linear programming problem. Further, the authors propose a distributed redundant constraint reduction scheme to cut down the number of constraints, and thus reduce the computational efforts of the problem. Although thematically \cite{icdcs14} is related to our current work, nevertheless our treatment of the subject of low radiation efficient charging is radically different. Indeed, this is due to the different charging model that we define, which takes into account hardware restrictions of the chargers and nodes of the network (energy and capacity bounds). These constraints introduce a non-linearity in our problems that did not appear in the treatment of \cite{icdcs14}.

\subsection{Our contribution} 

In this paper, we follow a new approach for radiation aware charging in wireless settings. In particular, as our first contribution in this paper, we define a \emph{new charging model} that greatly differs from existing models in that it takes into account hardware restrictions of the chargers and nodes of the network. More precisely, we assume (a) that chargers have \emph{finite initial energy supplies}, which restricts the amount of energy that they can transfer to nearby nodes and (b) that every node has \emph{finite battery capacity}, which restricts the total amount of energy that it can store. It is worth noting that previous works have only considered the problem of maximization of power (i.e. the rate of energy transfer) from the chargers to the nodes, thus ignoring such restrictions. However, new technological advances on Wireless Energy Transfer via Strongly Coupled Magnetic Resonances suggest that such restrictions are already in the heart of efficient energy management problems in such networks.

An important consequence of the energy and capacity restrictions in our model, which sets it apart from other models considered in the literature thus far, is that they introduce \emph{non-linear constraints} that radically change the nature of the computational problems we consider. In fact, our charging model implicitly introduces the notion of \emph{activity time} in the (radiation aware) charging process, which is the time that a wireless entity (i.e. charger or node) can affect the network. 

As our second contribution, we present and study the \emph{Low Radiation Efficient Charging Problem} (LREC). Rather than the maximization of the cumulative power on nodes, the objective function that we wish to optimize in LREC is the amount of ``useful'' energy transferred from chargers to nodes (under constraints on the maximum level of radiation caused because of the Wireless Energy Transfer). We present several fundamental properties of our objective function that highlight several obstacles that need to be overcome when studying LREC. Furthermore, we present an algorithm for computing the value of the objective function, given the configuration of the network at any time point, which runs in linear time in the number of chargers and nodes.   

As our third contribution, we present a relaxation of the LREC problem, namely the \emph{Low Radiation Disjoint Charging Problem} (RLDC), which simplifies the computation of the maximum electromagnetic radiation inside the area where chargers and nodes are deployed (i.e. the area of interest). We prove that even this seemingly easier version of our basic problem is NP-hard, by reduction from the Independent Set Problem in Disc Contact Graphs. Furthermore, we present an integer program for finding the optimal solution to RLDC. We approximately solve this integer program by using standard relaxation and rounding techniques and we use the computed (feasible) solution to assess the performance of our iterative heuristic solution to LREC.

In view of hardness indications for LREC, we propose an iterative local improvement heuristic \texttt{IterativeLREC}, which runs in polynomial time and we evaluate its performance via simulation. The most important feature of our algorithmic solution is that it decouples the computation of the objective function from the computation of the maximum radiation. Furthermore, our algorithmic solution is independent of the exact formula used for the computation of the point electromagnetic radiation. This is especially important, because due to the fact that the effect that multiple radiation sources have on the electromagnetic radiation is not well understood in our days. Finally, we provide extensive simulation results supporting our claims and theoretical results. We focus on three network metrics: \emph{charging efficiency}, \emph{maximum radiation} and \emph{energy balance}. 

%Note that the problem taken into account in this work, greatly differs from other problems in networks (e.g.~station management \cite{kranakis}), since we take into account wireless charging and radiation properties.

\section{The Model} \label{sec:Model}

We assume that there is a set of $n$ rechargeable \emph{nodes} ${\cal P} = \{v_1, v_2, \ldots, v_n\}$ and a set of $m$ wireless power \emph{chargers} ${\cal M} = \{u_1, u_2, \ldots, u_m\}$ which are deployed inside an area of interest ${\cal A}$ (say inside $\mathbb{R}^2$). Unless otherwise stated, we will assume that both nodes and chargers are static, i.e. their positions and operational parameters are specified at time 0 and remain unchanged from that time on. 

For each charger $u \in {\cal M}$, we denote by $E_u^{(t)}$ the \emph{available energy} of that charger, that it can use to charge nodes within some \emph{radius} $r_u$ (i.e. we assume that the initial energy of charger $u$ is $E_u^{(0)}$). The radius $r_u$ for each charger $u \in {\cal M}$, can be chosen by the charger at time 0 and remains unchanged for any subsequent time\footnote{The assumption of adjustable charging radius is broadly used in the Wireless Energy Transfer literature, e.g., \cite{icdcs14} and \cite{7218622}.} (hence the non-dependence of $r_u$ from $t$ in the notation). Furthermore, for each node $v \in {\cal P}$, we denote by $C_v^{(t)}$ the \emph{remaining energy storage capacity} of the node at time $t$ (i.e. the initial energy storage capacity of node $v$ is $C_v^{(0)}$). 

We consider the following well established \emph{charging model}\footnote{Relevant assumptions and models have been validated in the bibliography, e.g., \cite{suru1} and \cite{suru2}.}: a node $v \in {\cal P}$ harvests energy from a charger $u \in {\cal M}$ with \emph{charging rate} given by

\begin{equation} \label{eq-power}
P_{v, u}(t) = \left\{
\begin{array}{ll}
	\frac{\alpha  r_u^2}{(\beta + \textnormal{dist}(v, u))^2}, & \quad \textrm{if  $E_u^{(t)}, C_u^{(t)}>0$, $\textnormal{dist}(v, u) \leq r_u$} \\
	0, & \quad \textrm{otherwise.}
\end{array} \right.
\end{equation} 
$\alpha$ and $\beta$ are known positive constants determined by the environment and by hardware of the charger and the receiver. In particular, the above equation determines the rate at which a node $v$ harvests energy from any charger $u$ that has $v$ within its range, until the energy of $u$ is depleted or $v$ is fully charged. We stress out here that, besides its dependence on the geographic positions of $v$ and $u$, the charging rate $P_{v, u}(t)$ is also a function of time $t$. Indeed, for any node $v \in {\cal P}$ within distance $r_u$ from charger $u$, it is equal to $\frac{\alpha  r_u^2}{(\beta + \textnormal{dist}(v, u))^2}$ in a time-interval $[0, t^*_{u, v}]$ and 0 otherwise. By equation (\ref{eq-power}), the time point $t^*_{u, v}$ at which the value of $P_{v, u}(t)$ drops to 0 is the time when either the energy of $u$ is depleted or $v$ is fully charged (which also depends on other chargers that can reach $v$). Consequently, the exact value of $t^*_{u, v}$ may depend on the whole network (see also the discussion after Definition \ref{LREC} in Section \ref{sec:ProblemStatement}), i.e. the location, radius and initial energy of each charger and the location and initial energy storage capacity of each node. As a matter of fact, it seems that there is no ``nice'' closed formula for $t^*_{u, v}$. Nevertheless, the value of $t^*_{u, v}$ can be found by using the ideas of Section \ref{sec:ObjectiveFunction} and, more specifically, using a trivial modification of Algorithm \texttt{ObjectiveValue}.

Another crucial assumption on our charging model (which is also widely accepted by physicists) is that the harvested energy by the nodes is additive. Therefore, the total energy that node $v$ gets within the time interval $[0, T]$ is 

\begin{equation} \label{eq-harvest}
H_v(T) = \sum_{u \in {\cal M}} \int_0^{T} P_{v, u}(t) dt. 
\end{equation}

One of the consequences of our charging model (and in particular equations (\ref{eq-power}) and (\ref{eq-harvest}) above) is that $\sum_{u \in {\cal M}} E_u \geq \sum_{v \in {\cal P}} H_v(T)$, for any $T>0$. This means that the total energy harvested by the nodes cannot be larger than the total energy provided by the chargers. As yet another consequence, we have that $\sum_{v \in {\cal P}} C_v \geq \sum_{v \in {\cal P}} H_v(T)$, for any $T>0$, i.e. the total energy harvested by the nodes cannot be larger than the total energy that can be stored by all nodes.

To complete the definition of our model, we will make the assumption that the \emph{electromagnetic radiation (EMR)} at a point $x$ is proportional to the additive power received at that point. In particular, for any $x \in {\cal A}$, the EMR at time $t$ on $x$ is given by 

\begin{equation} \label{eq-radiation}
R_x(t) = \gamma \sum_{u \in {\cal M}} P_{x, u}(t), 
\end{equation}
where $\gamma$ is a constant that depends on the environment and $P_{x, u}(t)$ is given by equation (\ref{eq-power}). We note that, even though this is the usual assumption concerning electromagnetic radiation, the algorithmic solutions that we propose here could also be applied in the case of more general functions for $R_x(t)$ (as long as some quite general smoothness assumption are satisfied; see also Section \ref{sec:EMR}). We feel that this is especially important, because the notion of electromagnetic radiation is not completely understood in our days.

We finally note that, the existence of an energy (upper) bound for each charger and a capacity bound for each node greatly differentiates our model from other works in the literature. Indeed, not only can chargers decide on the length of their charging radius (a slight variation of which has been proposed in \cite{icdcs14}), but once each charger has made its decision, all chargers begin charging nodes within their radius until either their energy has been depleted, or every node within their radius has already reached its energy storage capacity. Furthermore, this characteristic radically changes the nature of the computational problem that we consider (see Section \ref{sec:ProblemStatement}).

\section{Problem Statement and First Results} \label{sec:ProblemStatement}

In general, we would like to use the chargers as efficiently as possible, but we would also like to keep radiation levels within acceptable levels. In particular, we are interested in the following computational problem which we refer to as Low Radiation Efficient Charging (LREC):

\begin{definition}[Low Radiation Efficient Charging (LREC)] \label{LREC}
Let ${\cal M}$ be a set of wireless power chargers and ${\cal P}$ be a set of rechargeable nodes which are deployed inside an area of interest ${\cal A}$. Suppose that each charger $u \in {\cal M}$ initially has available energy $E_u^{(0)}$, and each node $v \in {\cal P}$ has initial energy storage capacity $C_v^{(0)}$. Assign to each charger $u \in {\cal M}$ a radius $r_u$, so that the total usable energy given to the nodes of the network is maximized and the electromagnetic radiation at any point of ${\cal A}$ is at most $\rho$. We assume that all chargers start operating simultaneously at time 0 and charging follows the model described in Section \ref{sec:Model}.
\end{definition}

Let $\vec{r} = (r_u: u \in {\cal M}), \vec{E}^{(0)} = (E_u^{(0)}: u \in {\cal M})$ and $\vec{C}^{(0)} = (C_v^{(0)}: v \in {\cal P})$. In essence, the \emph{objective function} that we want to maximize in the LREC problem is the following: 

\begin{equation}
\begin{split}
f_{\textnormal{LREC}}\left(\vec{r}, \vec{E}^{(0)}, \vec{C}^{(0)} \right) &\stackrel{def}{=} \sum_{v \in {\cal P}} \left( \lim_{t \to \infty} C_v^{(t)} \right) \\&= \sum_{u \in {\cal M}} \left( E_u^{(0)} - \lim_{t \to \infty} E_u^{(t)} \right).
\end{split}
\end{equation}
The last equality follows from the fact that we are assuming loss-less energy transfer from the chargers to the nodes (obviously this easily extends to lossy energy tranfer, but we do not consider such models in this paper). In fact, we only need to consider finite values for $t$, because the energy values $E_u^{(t)}$ will be unchanged after time $t^* \stackrel{def}{=} \max_{v \in {\cal P}, u \in {\cal M}} t_{u, v}^*$, where $t^*_{u, v}$ is the time point at which the value of $P_{v, u}(t)$ drops to 0 (i.e. is the time when either the energy of $u$ is depleted or $v$ is fully charged). Therefore $f_{\textnormal{LREC}}\left(\vec{r}, \vec{E}^{(0)}, \vec{C}^{(0)} \right) = \sum_{v \in {\cal P}} C_v^{(t)} = \sum_{u \in {\cal M}} \left( E_u^{(0)} - E_u^{(t)} \right)$, for any $t \geq t^*$. The following lemma provides an upper bound on the value of $t^*$, which is independent of the radius choice for each charger.

\begin{lemma} \label{lemma-T^*}
$t^*$ can be at most $$T^* = \frac{(\beta + \max_{u \in {\cal M}, v \in {\cal P}} \textnormal{dist}(v, u))^2}{\alpha  (\min_{u \in {\cal M}, v \in {\cal P}} \textnormal{dist}(v, u))^2} \max_{u \in {\cal M}, v \in {\cal P}}\{E_{u}^{(0)}, C_{v}^{(0)}\}.$$
\end{lemma}
\proof Since $t^* \stackrel{def}{=} \max_{v \in {\cal P}, u \in {\cal M}} t_{u,v}^*$, we only need to provide an upper bound on $t_{u,v}^*$. To this end, without loss of generality, we assume that there is a charger $u_0$ and a node $v_0$ such that $u_0$ can reach $v_0$ (hence also $r_{u_0} \neq 0$) and $t^* = t_{u_0,v_0}^*$. Furthermore, we need to consider two cases, depending on whether $t_{u_0,v_0}^*$ is equal to (a) the time when the energy of the charger $u_0$ is depleted, or (b) the time when $v_0$ is fully charged. 

In case (a), by the maximality of $t_{u_0,v_0}^*$, we have that

\begin{eqnarray} \label{eq-case-a}
\begin{split}
E_{u_0}^{(0)} = \sum_{v \in {\cal P}} \int_0^{t_{u,v}^*} P_{v, u}(t) dt &\geq \int_0^{t_{u_0,v_0}^*} P_{v_0, u_0}(t) dt \\&= t_{u_0,v_0}^* \frac{\alpha  r_{u_0}^2}{(\beta + \textnormal{dist}(v_0, u_0))^2}
\end{split}
\end{eqnarray}
where in the first and last equality we used the fact that in case (a) $t_{u_0,v_0}^*$ is the time when the energy of the charger $u_0$ is depleted (hence $v_0$ has not yet exceeded its energy storage capacity).

In case (b), by equation (\ref{eq-harvest}) and the maximality of $t_{u_0,v_0}^*$, we have that 

\begin{eqnarray} \label{eq-case-b}
\begin{split}
C_{v_0}^{(0)} &= H_{v_0}(t_{u_0,v_0}^*) = \sum_{u \in {\cal M}} \int_0^{t_{u_0,v_0}^*} P_{v, u}(t) dt \\&\geq \int_0^{t_{u_0,v_0}^*} P_{v_0, u_0}(t) dt = t_{u_0,v_0}^* \frac{\alpha  r_{u_0}^2}{(\beta + \textnormal{dist}(v_0, u_0))^2}
\end{split}
\end{eqnarray}
where in the first and last equality we used the fact that in case (b) $t_{u_0,v_0}^*$ is the time when $v_0$ is fully charged (hence the energy of $u_0$ has not been depleted yet). 

By equations (\ref{eq-case-a}) and (\ref{eq-case-b}), we have that 

\begin{equation}
t_{u_0,v_0}^* \leq \max\{E_{u_0}^{(0)}, C_{v_0}^{(0)}\} \frac{(\beta + \textnormal{dist}(v_0, u_0))^2}{\alpha  r_{u_0}^2} 
\end{equation}
which completes the proof. \qed

It is worth noting here that, given $\vec{r}, \vec{E}^{(0)}$ and $\vec{C}^{(0)}$, the exact value of $f_{\textnormal{LREC}}\left(\vec{r}, \vec{E}^{(0)}, \vec{C}^{(0)} \right)$ can be computed by using Algorithm \texttt{ObjectiveValue} in Section \ref{sec:ObjectiveFunction}).

%In more technical terms, we would like to maximize $\sum_{u \in {\cal M}} \left[ E_u^{(0)} - \lim_{t \to \infty} E_u^{(t)} \right]$, subject to assuming the model described  

%\begin{eqnarray}
%\max \sum_{u \in {\cal M}} \left[ E_u^{(0)} - E_u^{(t^*)} \right] & \quad & \\
%\textnormal{s.t.} & \quad & \nonumber \\
%H_v(t^*) \leq C_v^{(0)}, & \quad & \forall v \in {\cal P}
%\end{eqnarray}

We now prove a Lemma that highlights some of the difficulties that we face when trying to find a solution to LREC. Furthermore, it sets LREC apart from other computational problems studied so far in the literature.

\begin{figure}[ht]
\centering 
\includegraphics[width=0.6\columnwidth]{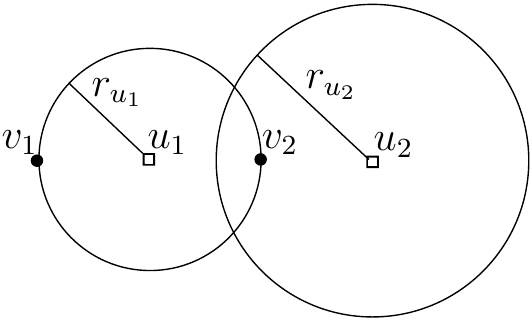} 
\caption{A network with 2 chargers $u_1, u_2$ and 2 nodes $v_1, v_2$. All 4 points are collinear and $\textnormal{dist}(v_1, u_1) = \textnormal{dist}(v_2, u_1) = \textnormal{dist}(v_2, u_2) = r_{u_1} = 1$.}
\label{ObjectiveFunctionCounterExample}
\end{figure}

\begin{lemma} \label{lemma-nonincreasing}
Let $\vec{r} = (r_u: u \in {\cal M}), \vec{E}^{(0)} = (E_u^{(0)}: u \in {\cal M})$ and $\vec{C}^{(0)} = (C_v^{(0)}: v \in {\cal P})$. The objective function $f_{\textnormal{LREC}}\left(\vec{r}, \vec{E}^{(0)}, \vec{C}^{(0)} \right)$ is not necessarily increasing in $\vec{r}$. Furthermore, the optimal radius for a charger is not necessarily equal to the distance from some node.
\end{lemma}
\proof Consider a network consisting of 2 chargers $u_1, u_2$ and 2 nodes $v_1, v_2$ all of which are collinear and $\textnormal{dist}(v_1, u_1) = \textnormal{dist}(v_2, u_1) = \textnormal{dist}(v_2, u_2) = r_{u_1} = 1$ (see Figure \ref{ObjectiveFunctionCounterExample}). Furthermore, assume for the sake of exposition of our arguments that $E_{u_1}^{(0)} = E_{u_2}^{(0)} = C_{v_1}^{(0)} = C_{v_2}^{(0)} = 1$. Finally, assume that the parameters for the charging rate in equation (\ref{eq-power}) are $\alpha=\beta=1$, the electromagnetic radiation parameter in equation (\ref{eq-radiation}) is $\gamma=1$ and that the upper bound on the radiation level is $\rho = 2$.

We will show that the optimal solution to the LREC problem in this network is when $r_{u_1} = 1$ and $r_{u_2} = \sqrt{2}$. To see this, first note that the electromagnetic radiation is maximum when $t=0$, i.e. when chargers are all operational. Furthermore, since there are only 2 radiation sources (namely chargers $u_1$ and $u_2$), it is not hard to verify that the electromagnetic radiation is maximized on the charger locations, i.e. $\max_{x, t} R_x(t) = \max\{r_{u_1}^2, r_{u_2}^2\}$. Consequently, since $\rho =2$, the radius of each charger can be at most $\sqrt{2}$. On the other hand, to achieve an objective function value that is larger than 1, both $r_{u_1}$ and $r_{u_2}$ must be at least 1. In fact, if $r_{u_1} = r_{u_2} \in [1, \sqrt{2}]$, then, by symmetry, $v_2$ will reach its energy storage capacity at the exact moment that the energy of $u_1$ is depleted. Therefore, the objective function value will be only $\frac{3}{2}$, since both $u_1$ and $u_2$ will have contributed the same amount of energy to fully charge $v_2$. To do better, $v_2$ must reach its energy storage capacity without using too much of the energy of $u_1$, which happens when $r_2 > r_1$. In this case, $v_2$ will reach its energy storage capacity before the energy of $u_1$ is depleted, and so $u_1$ will use the remaining energy to further charge $v_1$. In particular, when $r_2 > r_1$, we have that $t_{u_2,v_1}^* = t_{u_2,v_2}^*$ and also

\begin{equation}
1 = C_{v_2}^{(0)} = t_{u_2,v_1}^* \frac{r_{u_1}^2 + r_{u_2}^2}{4}.
\end{equation}  
The remaining energy of $u_1$ at time $t_{u_2,v_1}^*$ will then be 

\begin{equation}
E_{u_1}^{(0)} - 2 t_{u_2,v_1}^* \frac{r_{u_1}^2}{4} = 1 - 2\frac{r_{u_1}^2}{r_{u_1}^2 + r_{u_2}^2}.
\end{equation}
Since $1 \leq r_1 < r_2 \leq \sqrt{2}$, this is maximized for $r_{u_1} = 1$ and $r_{u_2} = \sqrt{2}$, in which case $E_{u_1}^{(t_{u_2,v_1}^*)} = \frac{1}{3}$. In that case, when $v_2$ reaches its energy storage capacity, $u_1$ will have $\frac{1}{3}$ energy units more to give solely to $v_1$; the other $\frac{2}{3}$ units will have been split evenly between $v_1$ and $v_2$. This means that the objective function value is $\frac{5}{3}$ and it is maximum.  

Now this example shows that, not only the radius of chargers are not necessarily equal to the distance from some node (since to achieve the optimum we must have $r_{u_2} = \sqrt{2}$), but also increasing $r_1$ will result in a sub-optimal objective function value.

\qed

\section{Computing the Objective Function} \label{sec:ObjectiveFunction}

In this section, we provide an algorithm for computing the value of our objective function (i.e. the amount of energy given by the chargers to the nodes), given the radii of the chargers, the capacities of the nodes and the available energies of the chargers. More precisely, assume that at some time $t$, each charger $u \in {\cal M}$ has remaining energy $E_u^{(t)}$ and each node $v \in {\cal P}$ can store $C_v^{(t)}$ energy. The tuple $\Sigma^{(t)} = (\vec{r}, \vec{E}^{(t)}, \vec{C}^{(t)})$, where $\vec{r} = (r_u: u \in {\cal M}), \vec{E}^{(t)} = (E_u^{(t)}: u \in {\cal M})$ and $\vec{C}^{(t)} = (C_v^{(t)}: v \in {\cal P})$, will be called the \emph{configuration of the network at time $t$}. For each $u \in {\cal M}$, we denote by ${\cal P}_u^{(t)} \stackrel{def}{=} \{v: \textnormal{dist}(v, u) \leq r_u, C_v^{(t)} > 0\}$ the set of nodes within distance $r_u$ from $u$ that have not reached their storage capacities at time $t$. Furthermore, for each $v \in {\cal P}$, we denote by ${\cal M}_v^{(t)} \stackrel{def}{=} \{u: v \in {\cal P}_u^{(t)}, E_u^{(t)} > 0\}$ the set of chargers that can reach $v$ and have not depleted their energy at time $t$. Finally, denote by ${\cal M}_{\emptyset}^{(t)} \stackrel{def}{=} \{u \in {\cal M}: E_u^{(t)} =0\}$ the set of chargers that have depleted their energy by time $t$. Similarly, denote by ${\cal P}_{\emptyset}^{(t)} \stackrel{def}{=} \{v \in {\cal P}: C_u^{(t)} =0\}$ the set of nodes that have reached their energy storage capacity by time $t$.

The value of the objective function can be computed by the following algorithm. The main idea is that, given the configuration of the network at any time $t$, we can find which will be the next charger (or node respectively) that will deplete his energy (resp. will reach its energy storage capacity) and when. The algorithm stops when no node can be charged any more, which happens either when they have reached their total capacity (i.e. $C_v^{(t)} = 0$), or all chargers that can reach it have depleted their energy (i.e. $\sum_{u \in {\cal M}_v^{(t)}} E_u^{(t)} = 0$).

\begin{algorithm}
\DontPrintSemicolon
\SetKwInOut{Input}{Input}\SetKwInOut{Output}{Output}
 \Input{Initial configuration $\Sigma^{(0)} = (\vec{r}, \vec{E}^{(0)}, \vec{C}^{(0)})$}
  Set $t = 0$
  
 \While{$\left[\bigcup_{v \in {\cal P}} \left\{ \left(C_v^{(t)} > 0 \right) \textnormal{ AND } \left(\sum_{u \in {\cal M}_v^{(t)}} E_u^{(t)} > 0 \right) \right\} \right]$}{
Let $t_{\cal M} = \min_{u \in {\cal M} \backslash {\cal M}_{\emptyset}^{(t)}}\{t': t' \sum_{v \in {\cal P}_u^{(t)}} P_{v, u}(t) = E_u^{(t)}\}$\;
Let $t_{\cal P} = \min_{v \in {\cal M} \backslash {\cal P}_{\emptyset}^{(t)}} \{t': t' \sum_{u \in {\cal M}_v^{(t)}} P_{v, u}(t) = C_v^{(t)} \}$\;
Let $t_0 = \min \{t_{\cal M}, t_{\cal P} \}$\;
For all $u \in {\cal M} \backslash {\cal M}_{\emptyset}^{(t)}$, set $E_u^{(t+t_0)} = E_u^{(t)} - t_0 \sum_{v \in {\cal P}_u^{(t)}} P_{v, u}(t)$\;
For all $v \in {\cal P} \backslash {\cal P}_{\emptyset}^{(t)}$, set $C_v^{(t+t_0)} = C_v^{(t)} - t_0 \sum_{u \in {\cal M}_v^{(t)}} P_{v, u}(t)$\;
Set $t = t+t_0$ and update ${\cal M}_{\emptyset}^{(t)}$ and ${\cal P}_{\emptyset}^{(t)}$\;
}
 
\Output{$\sum_{u \in {\cal M}} (E_u^{(0)} - E_u^{(t)})$}
 \caption{\texttt{ObjectiveValue}}
\end{algorithm}

Notice that, in every iteration, algorithm \texttt{ObjectiveValue} sets to 0 the energy level or the capacity of at least one charger or node. Therefore, we have the following:

\begin{lemma} \label{lemma-objectivevaluetime}
Algorithm \texttt{ObjectiveValue} terminates in at most $n+m$ while-iterations.
\end{lemma}

\section{Computing the Maximum Radiation} \label{sec:EMR}

One of the challenges that arises in our model is the computation of the maximum radiation inside the area of interest ${\cal A}$, as well as the point (or points) where this maximum is achieved. Unfortunately, it is not obvious where the maximum radiation is attained inside our area of interest and it seems that some kind of discretization is necessary. In fact, in our experiments, we use the following generic MCMC procedure: for sufficiently large $K \in \mathbb{N}^+$, choose $K$ points uniformly at random inside ${\cal A}$ and return the maximum radiation among those points. We note also that, the computation of the electromagnetic radiation at any point takes $O(m)$ time, since it depends only on the distance of that point from each charger in ${\cal M}$.

One of the main drawbacks of the above method for computing the maximum radiation is that the approximation it achieves depends on the value of $K$ (which is equivalent to how refined our discretization is). On the other hand, it does not take into account the special form of the electromagnetic radiation in equation (\ref{eq-radiation}). In fact, our iterative algorithm \texttt{IterativeLREC} in Section \ref{sec:HeuristicLREC} does not depend on the specific form of equation (\ref{eq-radiation}), and this could be desirable in some cases (especially since the effect that multiple radiation sources have on the electromagnetic radiation is not completely understood).

%Nevertheless, we can find a better approximation to the maximum radiation by doing the following:

\section{A Local Improvement Heuristic for LREC.} \label{sec:HeuristicLREC}

We now present a heuristic for approximating the optimal solution to LREC. To this end, we first note that, for any charger $u \in {\cal M}$, we can approximately determine the radius $r_u$ of $u$ that achieves the best objective function value, given the radii $\vec{r}_{-u} = (r_{u'}: u' \in {\cal M} \backslash u)$ as follows:
Let $r_u^{\textnormal{max}}$ be the maximum distance of any point in ${\cal A}$ from $u$ and let $l \in \mathbb{N}^+$ be a sufficiently large integer. For $i = 0, 1, \ldots, l$, set $r_u = \frac{i}{l} r_u^{\textnormal{max}}$ and compute the objective function value (using algorithm \texttt{ObjectiveValue}) as well as the maximum radiation (using the method described in Section \ref{sec:EMR}). Assign to $u$ the radius that achieves the highest objective function value that satisfies the radiation constraints of LREC. Given that the discretization of ${\cal A}$ used to compute the maximum radiation has $K$ points in it, and using Lemma \ref{lemma-objectivevaluetime}, we can see that the number of steps needed to approximately determine the radius $r_u$ of $u$ using the above procedure is $O\left((n+m)l + mK \right)$. It is worth noting that we could generalize the above procedure to any number $c$ of chargers, in which case the running time would be $O\left((n+m)l^c + mK \right)$. In fact, for $c=m$ we would have an exhaustive-search algorithm for LREC, but the running time would be exponential in $m$, making this solution impractical even for a small number of chargers.

The main idea of our heuristic \texttt{IterativeLREC} is the following: in every step, choose a charger $u$ uniformly at random and find (an approximation to) the optimal radius for $u$ given that the radii of all other chargers are fixed. To avoid infinite loops, we stop the algorithm after a predefined number of iterations $K' \in \mathbb{N}^+$.

\begin{algorithm}[t!]
\DontPrintSemicolon
\SetKwInOut{Input}{Input}\SetKwInOut{Output}{Output}
 \Input{Charger and node locations}
$counter = 1$\;
\Repeat{$counter =K'$}{
Select u.a.r.~a charger $u \in {\cal M}$\;
Find (an approximation to) the optimal radius for $u$ given that the radii of all other chargers are fixed\;
$counter = counter + 1$\;
}
 
\Output{$\vec{r} = (r_u: u \in {\cal M})$}
 \caption{\texttt{IterativeLREC}}
\end{algorithm}

By the above discussion, \texttt{IterativeLREC} terminates in $O\left(K'(nl+ml + mK) \right)$ steps.

\section{A Relaxation of LREC} \label{section:radiation-oriented}

The intractability of the LREC problem is mainly due to the following reasons: (a) First, there is no obvious closed formula for the maximum radiation inside the area of interest ${\cal A}$ as a function of the positions and the radii of the chargers. (b) Second, as is suggested by Lemma \ref{lemma-nonincreasing}, there is no obvious potential function that can be used to identify directions inside $\mathbb{R}^m$ that can increase the value of our objective function.

In this section we consider the following relaxation to the LREC problem, which circumvents the problem of finding the maximum radiation caused by multiple sources:

\begin{definition}[Low Radiation Disjoint Charging (LRDC)] \label{LRDC}
Let ${\cal M}$ be a set of wireless power chargers and ${\cal P}$ be a set of rechargeable nodes which are deployed inside an area of interest ${\cal A}$. Suppose that each charger $u \in {\cal M}$ initially has available energy $E_u^{(0)}$, and each node $v \in {\cal P}$ has initial energy storage capacity $C_v^{(0)}$. Assign to each charger $u \in {\cal M}$ a radius $r_u$, so that the total usable energy given to the nodes of the network is maximized and the electromagnetic radiation at any point of ${\cal A}$ is at most $\rho$. We assume that all chargers start operating simultaneously at time 0 and that charging follows the model described in Section \ref{sec:Model}. Additionally, we impose the constraint that no node should be charged by more than 1 charger.
\end{definition}

The following Theorem concerns the hardness of LRDC.

\begin{theorem}
LRDC is NP-hard.
\end{theorem}
\proof The hardness follows by reduction from the Independent Set in Disc Contact Graphs \cite{garey76}. Let $G$ be a disc contact graph, i.e. a graph where vertices correspond to discs any two of which have at most 1 point in common. In particular, the set of vertices of $G$ corresponds to a set of $m$ discs $D(u_1, r_1), D(u_2, r_2), \ldots, D(u_m, r_m)$, where $D(u_j, r_j)$ is a disc centered at $u_j$ with radius $r_j$. Two vertices of $G$ are joined by an edge if and only if their corresponding discs have a point in common.

We now construct an instance of the LRDC as follows: We place a node on each disc contact point and, for $j = 1, 2, \ldots, m$, let $k_j$ be the maximum number of nodes in the circumference of the disc $D(u_j, r_j)$. We then add nodes on the circumference of every other disc in such a way that every disc has exactly the same number of nodes (say) $K$ uniformly around its circumference (notice that this is possible since every disc shares at most $m$ points of its circumference with other discs). We now place a charger on the center of each disc and set the radius bound for the charger corresponding to $u_j$ equal to $r_j$, for every $j = 1, 2, \ldots, m$. Finally, we set the initial energy storage capacity of each node equal to 1, the available energy of each charger equal to $K$ and the electromagnetic radiation bound $\rho = \max_{j \in [m]} \frac{\alpha r_j^2}{\beta^2}$. 

It is now evident that an optimal solution to LRDC on the above instance yields a maximum independent set in $G$; just pick disc $D(u_j, r_j)$ if the $j$-th charger has radius equal to $r_j$ and discard it otherwise.

\qed

We now present an integer program formulation for LRDC (to which we refer as \texttt{IP-LRDC}). To this end, we first note that, for any charger $u \in {\cal M}$, the distance of nodes/points in ${\cal P}$ from $u$ defines a (complete) ordering $\sigma_u$ in ${\cal P}$. In particular, for any two nodes $v, v' \in {\cal P}$ and a charger $u \in {\cal M}$, we will write $v \leq_{\sigma_u} v'$ if and only if $\textnormal{dist}(v, u) \leq \textnormal{dist}(v, u)$. For any charger $u$, define $i^{(u)}_{\textnormal{rad}}$ to be the furthest node from $u$ that can be charged by $u$ without $u$ violating the radiation threshold $\rho$ on its own. Similarly, define $i^{(u)}_{\textnormal{nrg}}$ to be the furthest node from $u$ with the property that if $u$ has radius at least $\textnormal{dist}(i^{(u)}_{\textnormal{nrg}}, u)$, then the energy of $u$ will be fully spent. Assuming we break in $\sigma$ arbitrarily, nodes $i^{(u)}_{\textnormal{rad}}$ and $i^{(u)}_{\textnormal{nrg}}$ are uniquely defined for any charger $u$. Our integer program solution is presented below. 

\begin{equation}
\max \sum_{u \in {\cal M}} \left(E^{(0)}_u x_{i^{(u)}_{\textnormal{nrg}}, u} + \sum_{v \leq_{\sigma_u} i^{(u)}_{\textnormal{nrg}}} (x_{v, u} - x_{i^{(u)}_{\textnormal{nrg}}, u}) C_v^{(0)} \right)
\end{equation}

\vspace{-0.3cm}

\begin{eqnarray}
\textnormal{subject to:} & \quad & \nonumber \\
\sum_{u \in {\cal M}} x_{v, u} \leq 1, & \quad & \forall v \in {\cal P} \label{constraint-chargeuniqueness} \\
x_{v, u} - x_{v', u} \geq 0, & \quad & \forall v, v' \in {\cal P}, \forall u \in {\cal M}: \nonumber\\
& \quad &
v \leq_{\sigma_u} v' \label{constraint-increasing} \\
x_{v, u} = 0, & \quad & \forall v \in {\cal P}, \forall u \in {\cal M}: \nonumber\\
& \quad &
v >_{\sigma_u} i^{(u)}_{\textnormal{rad}} \textnormal{ or } v >_{\sigma_u} i^{(u)}_{\textnormal{nrg}} \label{constraint-logicalbound} \\
x_{v, u} \in \{0, 1\}, & \quad & \forall v \in {\cal P}, \forall u \in {\cal M}. 
\end{eqnarray}

In \texttt{IP-LRDC}, variable $x_{v, u}$ indicates whether or not the (unique) charger that can reach $v$ is $u$. The existence of at most one charger per node in a feasible assignment of LRDC is guaranteed by constraint (\ref{constraint-chargeuniqueness}). Constraint (\ref{constraint-increasing}) guarantees that when a node $v'$ can be reached by $u$, then all nodes closer to $u$ can also be reached by $u$. Finally, constraint (\ref{constraint-logicalbound}) guarantees that the radiation threshold is not violated and also suggests that there is no reason why a charger should be able to reach nodes that are further than $i^{(u)}_{\textnormal{nrg}}$.

To understand the objective function that we want to maximize in \texttt{IP-LRDC}, notice that, for any charger $u \in {\cal M}$, if $r_u \geq \textnormal{dist}(i^{(u)}_{\textnormal{nrg}}, u)$ (which is equivalent to having $x_{i^{(u)}_{\textnormal{nrg}}, u} =1$), then the useful energy transferred from $u$ to the nodes of the network will be exactly $E^{(0)}_u$. Indeed, this is captured by our objective function, since $E^{(0)}_u x_{i^{(u)}_{\textnormal{nrg}}, u} + \sum_{v \leq_{\sigma_u} i^{(u)}_{\textnormal{nrg}}} (x_{v, u} - x_{i^{(u)}_{\textnormal{nrg}}, u}) C_v^{(0)} = E^{(0)}_u$, when $x_{i^{(u)}_{\textnormal{nrg}}, u} =1$, since, by constraint (\ref{constraint-increasing}), we have that $x_{v, u} = x_{i^{(u)}_{\textnormal{nrg}}, u}$, for any $v \leq_{\sigma_u} i^{(u)}_{\textnormal{nrg}}$. On the other hand, when $x_{i^{(u)}_{\textnormal{nrg}}, u} = 0$, charger $u$ will not be able to spend all of its energy, since the nodes it can reach cannot store all of it. This is also captured by our objective function, since $E^{(0)}_u x_{i^{(u)}_{\textnormal{nrg}}, u} + \sum_{v \leq_{\sigma_u} i^{(u)}_{\textnormal{nrg}}} (x_{v, u} - x_{i^{(u)}_{\textnormal{nrg}}, u}) C_v^{(0)} = \sum_{v \leq_{\sigma_u} i^{(u)}_{\textnormal{nrg}}} x_{v, u} C_v^{(0)}$, when $x_{i^{(u)}_{\textnormal{nrg}}, u} =0$, which is equal to the total energy that the nodes reachable from $u$ could harvest in total. 

In our experimental evaluation, we solve \texttt{IP-LRDC} by first making a linear relaxation and then rounding the solution so that the constraints (\ref{constraint-chargeuniqueness}), (\ref{constraint-increasing}) and (\ref{constraint-logicalbound}). It is easy to see that the objective function value that we get is a lower bound on the optimal solution of the LREC problem. We use this bound to evaluate the performance of our iterative algorithm \texttt{IterativeLREC} (see Section \ref{sec:HeuristicLREC}).

\begin{figure*}[t!]
        \centering
        \begin{subfigure}[b]{0.33\textwidth}%{\columnwidth}
                \includegraphics[width=\textwidth]{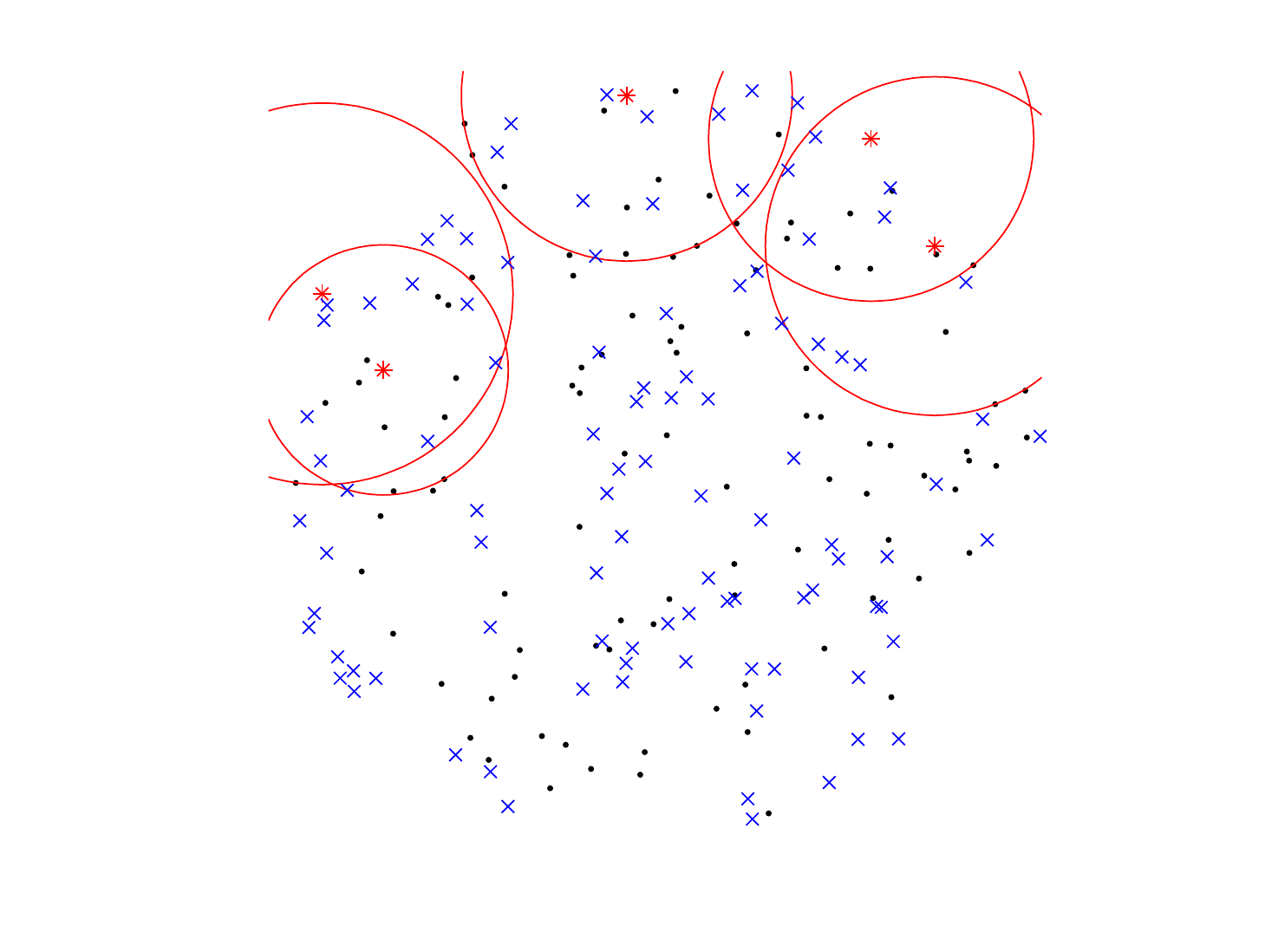}
                \caption{\texttt{ChargingOriented}}
                \label{fig:}
        \end{subfigure}%
        ~ %add desired spacing between images, e. g. ~, \quad, \qquad etc.
        \begin{subfigure}[b]{0.33\textwidth}%{\columnwidth}
                \includegraphics[width=\textwidth]{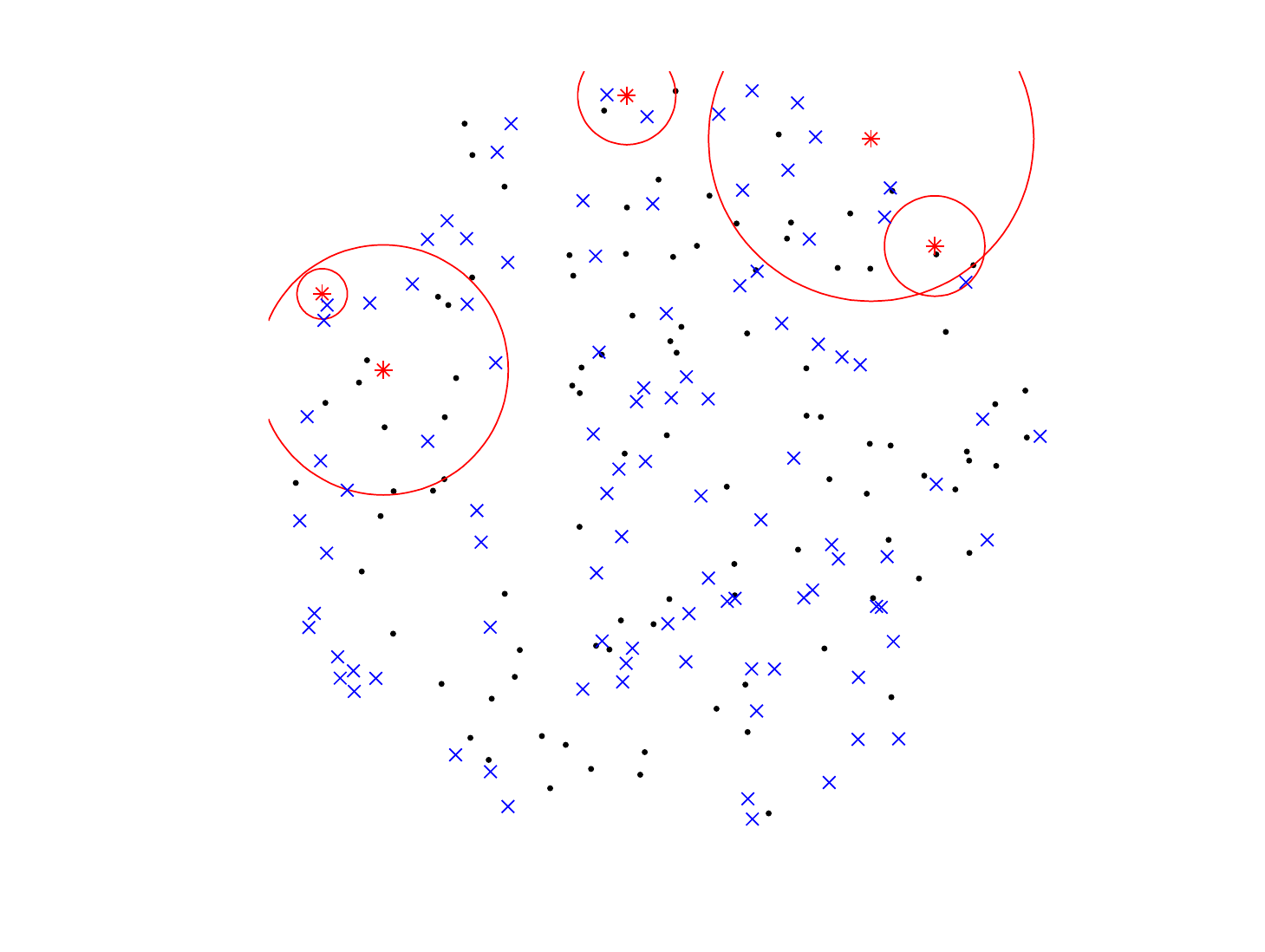}
                \caption{\texttt{IterativeLREC}}
                \label{fig:}
        \end{subfigure}%
        ~ %add desired spacing between images, e. g. ~, \quad, \qquad etc.
        \begin{subfigure}[b]{0.33\textwidth}%{\columnwidth}
                \includegraphics[width=\textwidth]{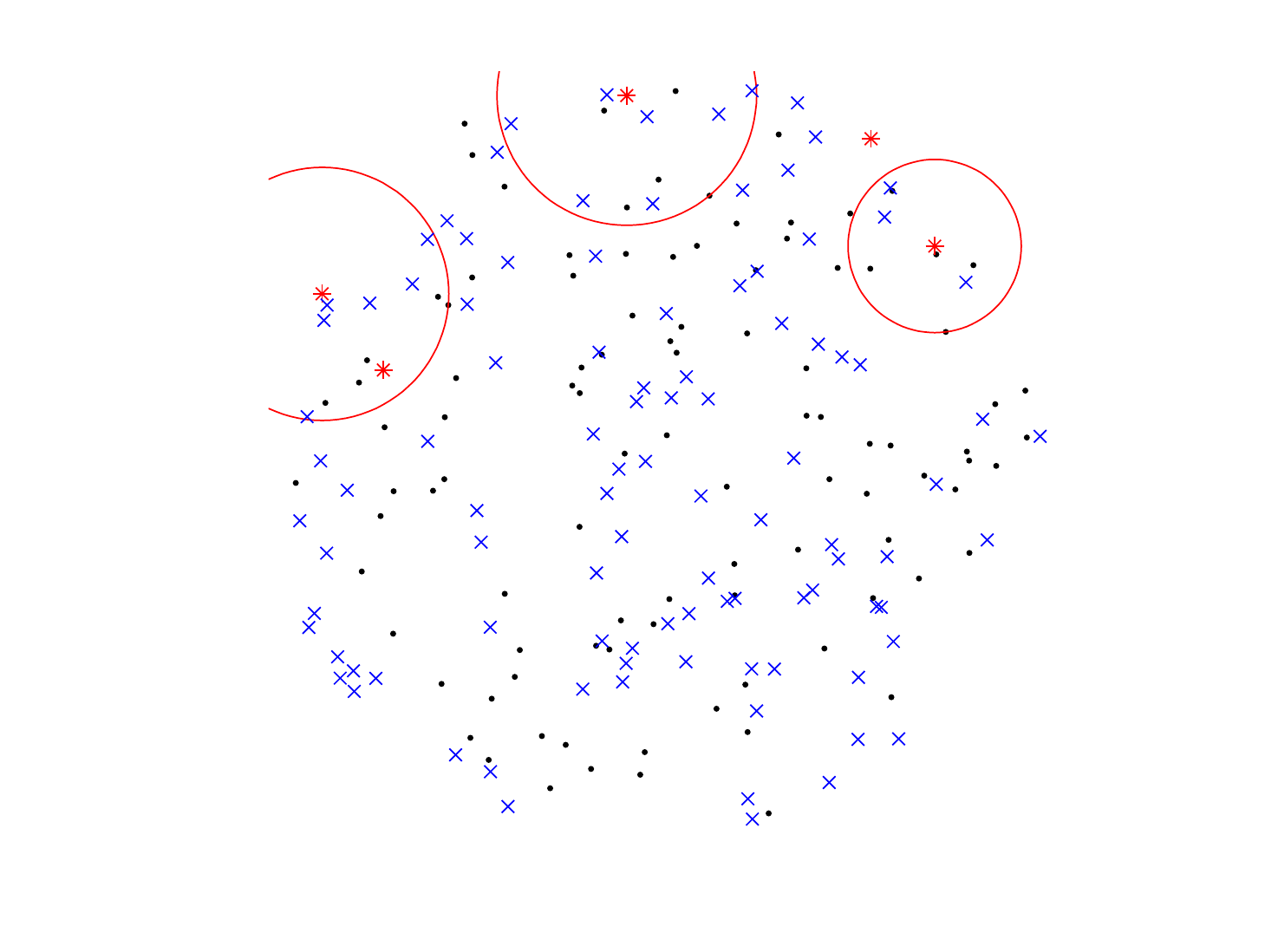}
                \caption{\texttt{IP-LRDC}}
                \label{fig:}
        \end{subfigure}%
        \caption{Network snapshot using 5 chargers.}
        \label{fig:snapshot}
        \vspace{-0.5cm}
\end{figure*}

\begin{figure*}[t!]
\centering
        \begin{subfigure}[b]{0.8\columnwidth}
                \includegraphics[width=\textwidth]{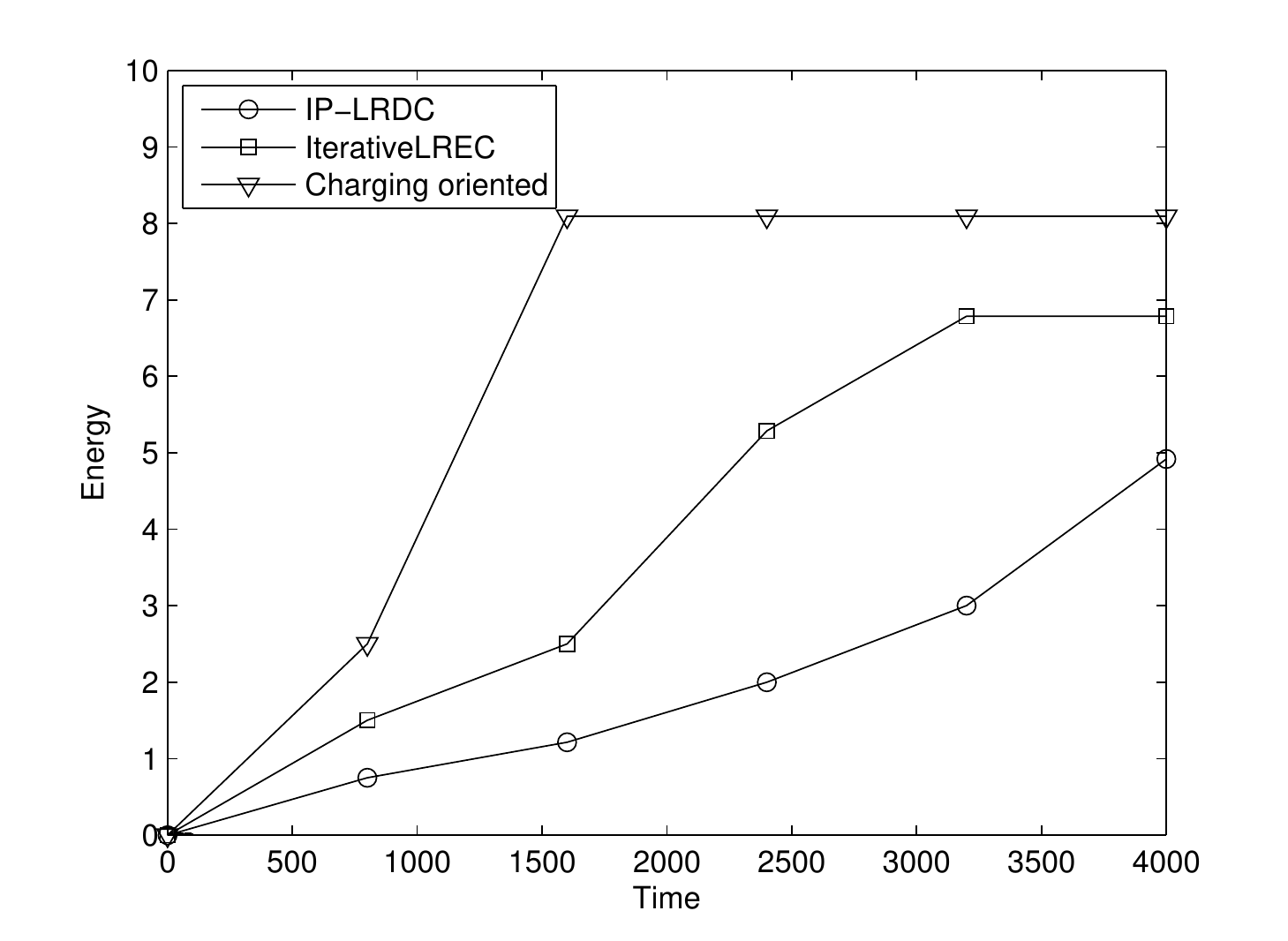}
                \caption{Charging efficiency over time.}
        	\label{fig:efficiency}
        \end{subfigure}%
        \begin{subfigure}[b]{0.8\columnwidth}
                \includegraphics[width=\textwidth]{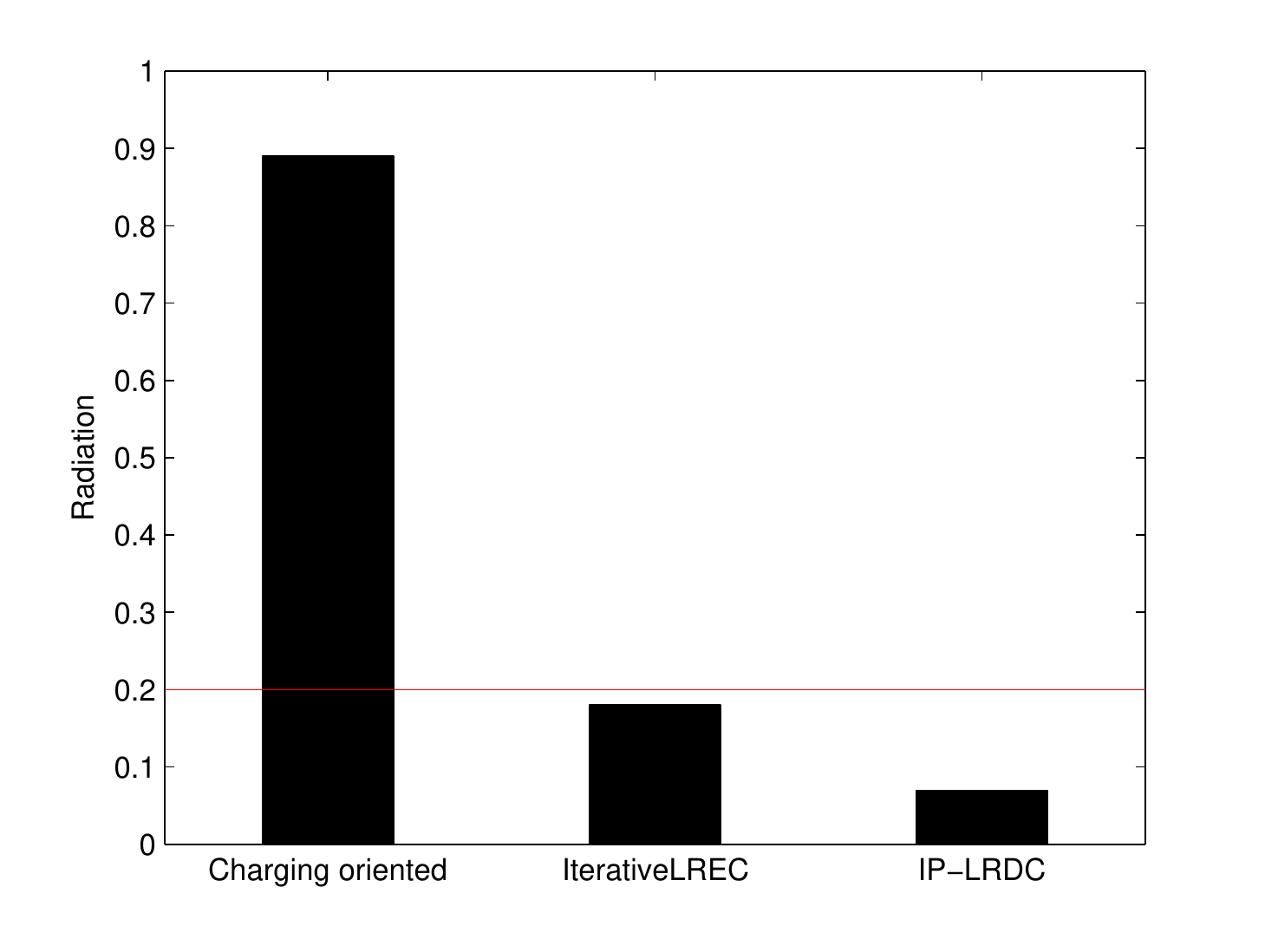}
                \caption{Maximum radiation.}
        \label{fig:radiation}
        \end{subfigure}%
        \caption{Efficiency and radiation.}
        \vspace{-0.5cm}
\end{figure*}

\begin{comment}

\begin{figure}[t!]
        \begin{subfigure}[b]{\columnwidth}
                \includegraphics[width=\textwidth]{chargers1-eps-converted-to.pdf}
                \caption{Different numbers of chargers.}
                \label{fig:}
        \end{subfigure}%
        
        \begin{subfigure}[b]{\columnwidth}
                \includegraphics[width=\textwidth]{chargers2-eps-converted-to.pdf}
                \caption{Maximum radiation.}
                \label{fig:}
        \end{subfigure}%
        \caption{Various parameters.}
\end{figure}
\end{comment}

\begin{figure*}[t!]
        \centering
        \begin{subfigure}[b]{0.32\textwidth}
                \includegraphics[width=\textwidth]{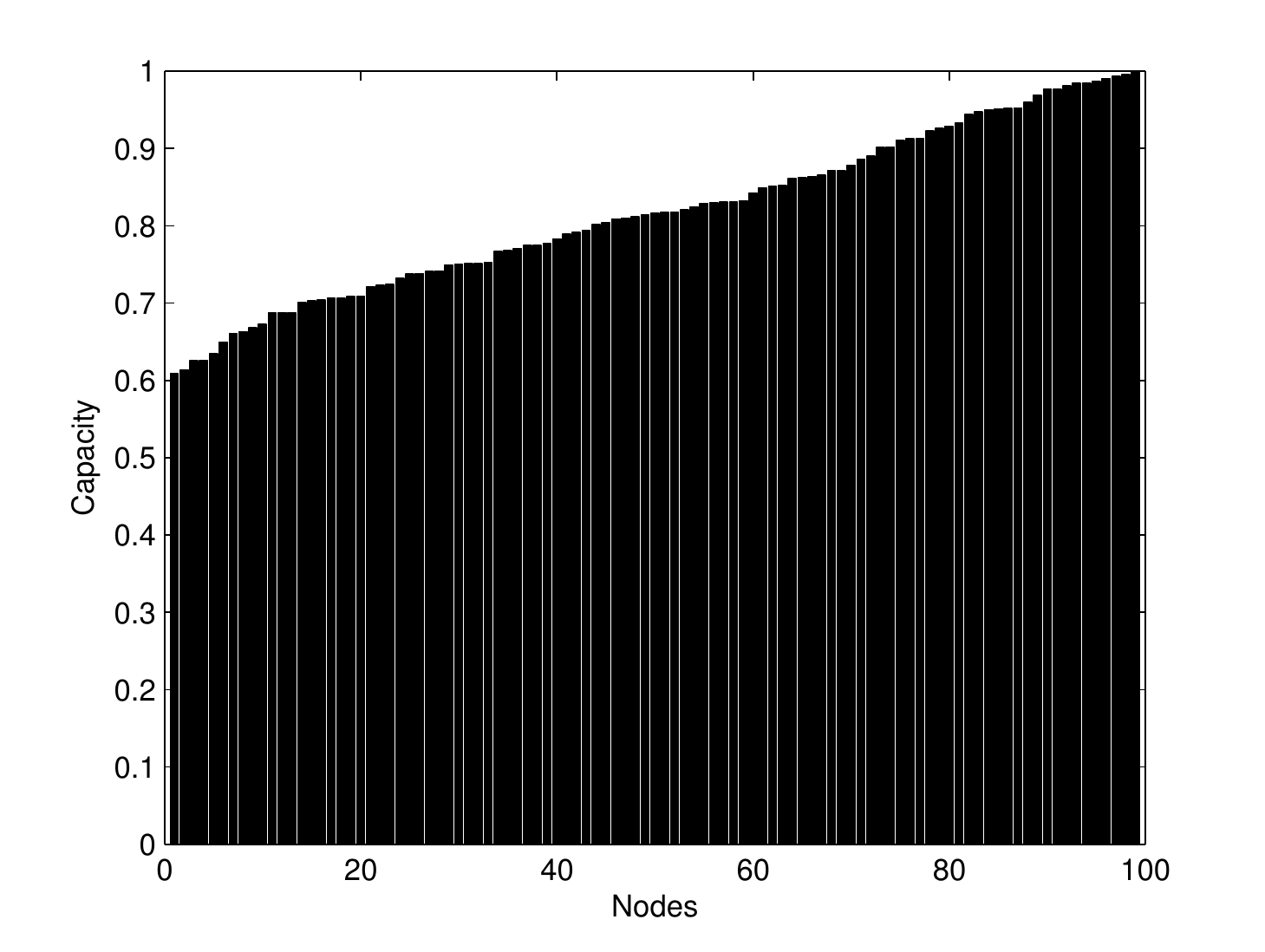}
                \caption{\texttt{ChargingOriented}}
                \label{fig:}
        \end{subfigure}%
        ~ %add desired spacing between images, e. g. ~, \quad, \qquad etc.
        \begin{subfigure}[b]{0.32\textwidth}
                \includegraphics[width=\textwidth]{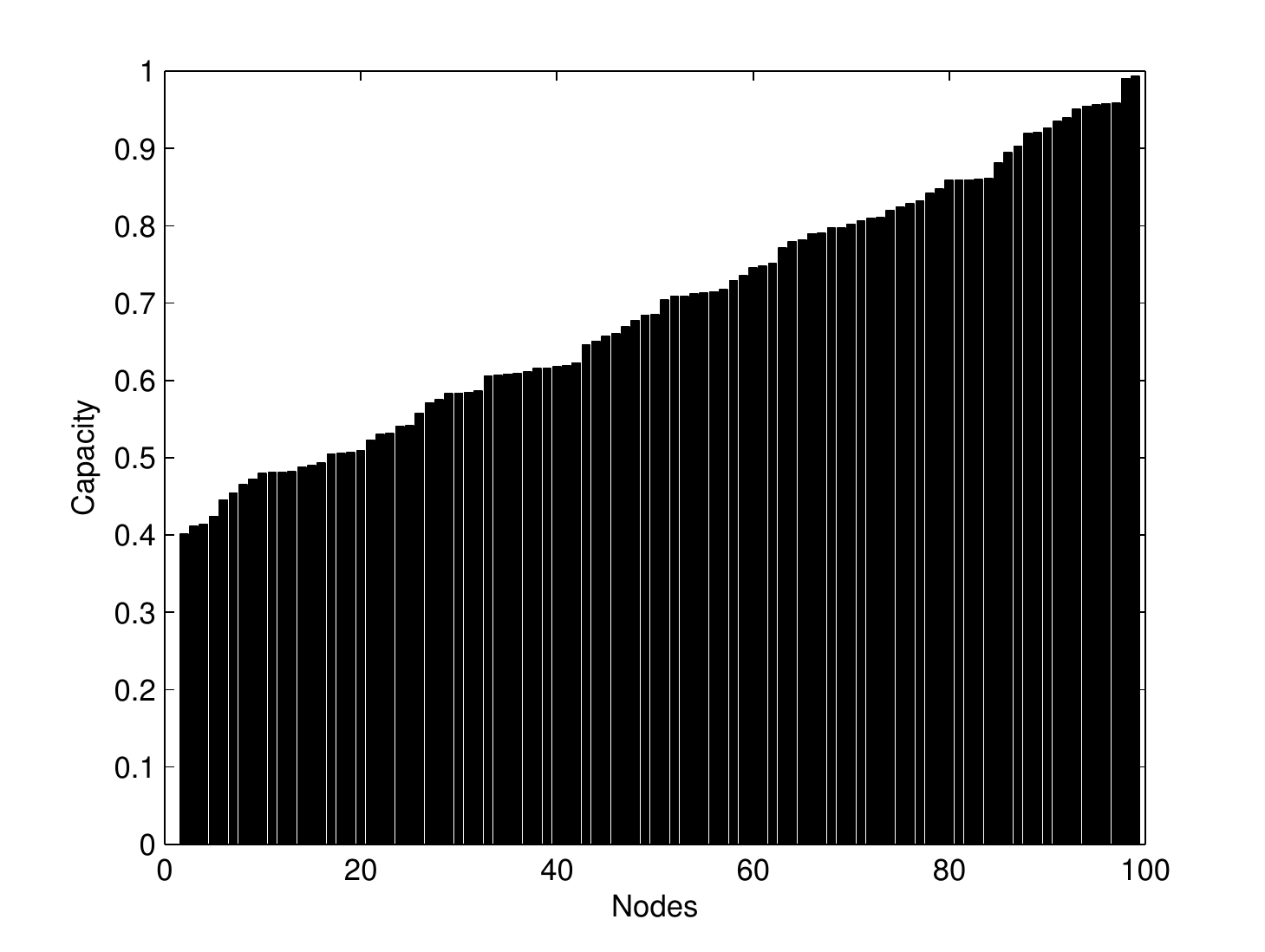}
                \caption{\texttt{IterativeLREC}}
                \label{fig:}
        \end{subfigure}%
        ~ %add desired spacing between images, e. g. ~, \quad, \qquad etc.
        \begin{subfigure}[b]{0.32\textwidth}
                \includegraphics[width=\textwidth]{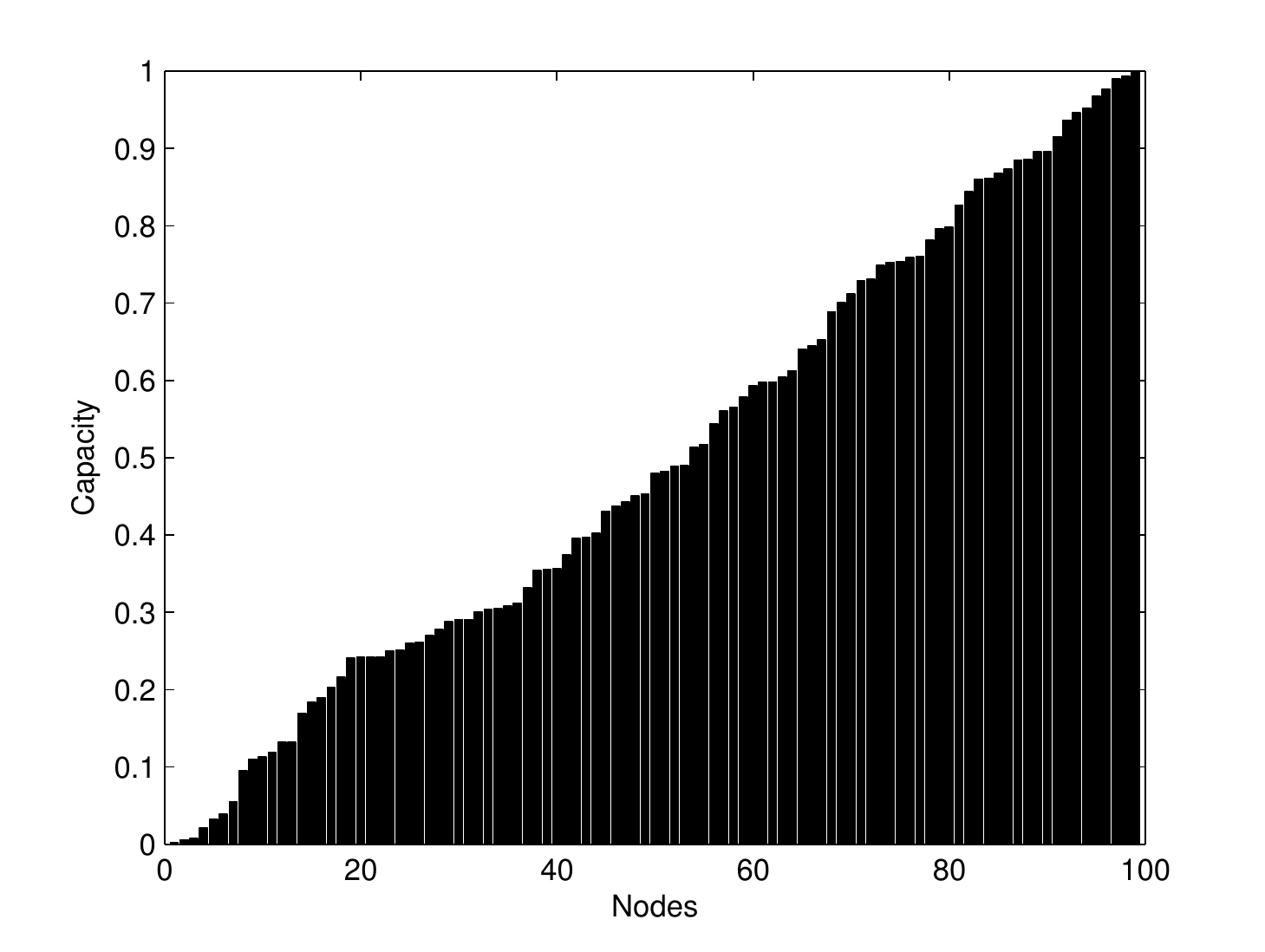}
                \caption{\texttt{IP-LRDC}}
                \label{fig:}
        \end{subfigure}%
        \caption{Energy balance.}
        \label{fig:balance}
        \vspace{-0.5cm}
\end{figure*}

\section{Performance Evaluation}

We conducted simulations in order to evaluate our methods using Matlab R2014b. We compared \texttt{IterativeLREC}, \texttt{IP-LRDC} (after the linear relaxation) and a charger configuration scheme in which each charger $u$ sets its radius equal to $\text{dist}(u,i_{rad}^{(u)})$. We call this new configuration ``\texttt{ChargingOriented}'' because it assigns the maximum radius to each charger, without individually violating the radiation threshold. In other words, this configuration provides the best possible rate of transferring energy in the network and serves as an upper bound on the charging efficiency of the performance of \texttt{IterativeLREC}, but is expected to achieve a poor performance on keeping the radiation low, due to frequent, large overlaps. A snapshot of a uniform network deployment with $|\mathcal{P}|=100, |\mathcal{M}|=5$ and $K=100$, is shown in Fig.~\ref{fig:snapshot}. We observe that the radii of the chargers in the \texttt{ChargingOriented} case are larger than the other two cases. In the case of \texttt{IP-LRDC} the radiation constraints lead to a configuration where two chargers are not operational. \texttt{IterativeLREC} provides a configuration in between the \texttt{ChargingOriented} and \texttt{IP-LRDC}, in which some overlaps of smaller size are present.

We deploy uniformly at random %in a $50 \times 50$ area, 
$|\mathcal{P}|=100$ network nodes of identical capacity, $|\mathcal{M}|=10$ wireless chargers of identical energy supplies and $K=1000$ points of radiation computation. We set $\alpha = 0, \beta = 1, \gamma = 0.1$ and $\rho = 0.2$. For statistical smoothness, we apply the deployment of nodes in the network and repeat each experiment 100 times. The statistical analysis of the findings (the median, lower and upper quartiles, outliers of the samples) demonstrate very high concentration around the mean, so in the following figures we only depict average values. We focus our simulations on three basic metrics: charging efficiency, maximum radiation and energy balance.

\textbf{Charging efficiency.} The objective value that is achieved as well as the time that is spent for the charging procedure is of great importance to us. The objective values achieved were $80.91$ by the \texttt{ChargingOriented}, $67.86$ by the \texttt{IterativeLREC} and $49.18$ by the \texttt{IP-LRDC}. The \texttt{ChargingOriented} method is the most efficient and quick, as expected but it results in high maximum radiation. As we observe in Fig.~\ref{fig:efficiency}, it distributed the energy in the network in a very short time. The efficiency of \texttt{ChargingOriented} both in terms of objective value and in terms of time is explained by the frequent charger radii overlaps that are created during the configuration of the chargers (e.g., Fig.~\ref{fig:snapshot}). \texttt{IP-LRDC} achieves the lowest efficiency of all due to the small charging radii and consequently small network coverage by the chargers. Our heuristic \texttt{IterativeLREC} achieves high enough efficiency w.r.t.~the radiation constraints. It's performance lies between the performance of \texttt{ChargingOriented} and \texttt{IP-LRDC}, both in terms of objective value and in terms of time.

\textbf{Maximum radiation.} The maximum amount of radiation incurred is very important regarding the safety impact of the corresponding charging method. High amounts of radiation, concentrated in network regions may render a method non-practical for realistic applications. This is the case for the \texttt{ChargingOriented}, which in spite of being very (charging) efficient, it significantly violates the radiation threshold (Fig.~\ref{fig:radiation}). \texttt{IterativeLREC} is performing very well, since it does not violate the threshold but in the same time provides the network with high amount of energy.

\textbf{Energy balance.} The energy balance property is crucial for the lifetime of Ad hoc Networks, since early disconnections are avoided and nodes tend to save energy and keep the network functional for as long as possible. For this reason, apart from achieving high charging efficiency, an alternative goal of a charging method is the balanced energy distribution among the network nodes. Fig.~\ref{fig:balance} is a graphical depiction of the energy provided in the network throughout the experiment. The nodes are sorted by their final energy level and by observing the Figure, we are able to make conclusions about the objective value and the energy balance of each method. Our \texttt{IterativeLREC} achieves efficient energy balance that approximates the performance of the powerful \texttt{ChargingOriented}.

\section{Conclusion}

In this paper, we define a new charging model and we present and study the Low Radiation Efficient Charging Problem, in which we wish to optimize the amount of ``useful'' energy transferred from chargers to nodes (under constraints on the maximum level of radiation). We present several fundamental properties of this problem and provide indications of its hardness. Also, we propose an iterative local improvement heuristic for LREC, which runs in polynomial time and we evaluate its performance via simulation. Our algorithm decouples the computation of the objective function from the computation of the maximum radiation and also does not depend on the exact formula used for the computation of the point electromagnetic radiation. We provide extensive simulation results supporting our claims and theoretical results. As future work, the design, as well as the theoretical and experimental evaluation of approximation algorithms with performance guarantees would be an interesting and meaningful research direction.



%\bibliographystyle{IEEEtran}

%\section*{References}

\balance
%\bibliographystyle{elsarticle-num}
%\bibliography{icdcs}

\end{document}